\documentclass{article}

\usepackage{amsfonts}
\usepackage{amssymb}
\usepackage{amsmath}
\usepackage{amsthm}

\usepackage{graphics}
\usepackage{subfigure}
\usepackage{epsfig}
\usepackage{graphicx}
\DeclareGraphicsExtensions{.eps, .ps}

\usepackage{multirow}

\numberwithin{equation}{section}

\begin{document}

\title{On the use of stabilising transformations for detecting
unstable periodic orbits in the Kuramoto-Sivashinsky equation}
\author{Jonathan J. Crofts and Ruslan L. Davidchack
        \footnote{Corresponding author: e-mail: {\sf r.davidchack@mcs.le.ac.uk},
        Phone: +44(0)\,116\,252\,3819, Fax: +44(0)\,116\,252\,3915}}

\maketitle

\begin{abstract}

In this paper we develop further a method for detecting unstable periodic orbits (UPOs) by stabilising transformations,
where the strategy is to transform the system of interest in such a way that the orbits become stable. The main
difficulty of using this method is that the number of transformations, which were used in the past, becomes overwhelming as
we move to higher dimensions~\cite{Lai99,Schmelcher97,Schmelcher98}. We have recently proposed a set of stabilising
transformations which is constructed from a small set of already found UPOs~\cite{Crofts06}. The main benefit of using the
proposed set is that its cardinality depends on the dimension of the unstable manifold at the UPO rather than the dimension
of the system. In a typical situation the dimension of the unstable manifold is much smaller than the dimension of the
system so the number of transformations is much smaller. Here we extend this approach to high-dimensional systems of ODEs
and apply it to the model example of a chaotic spatially extended system -- the Kuramoto-Sivashinsky equation. A comparison
is made between the performance of this new method against the competing methods of Newton-Armijo (NA) and
Levernberg-Marquardt (LM). In the latter case, we take advantage of the fact that the LM algorithm is able to solve
under-determined systems of equations, thus eliminating the need for any additional constraints.

\end{abstract}

\section{Introduction}

The following work is concerned with the detection of UPOs for large systems of ODEs of the form
\begin{equation}\label{eqn:ode}
\frac{dx}{dt} = v(x),
\end{equation}
with $x\in\mathbb{R}^n$ and $v(x)\in\mathbb{R}^n$. Often Eq.~(\ref{eqn:ode}) is the result of a spatial discretisation
of a parabolic PDE. In this way dynamical system results can be applied to extended systems which exhibit chaos. Over the
past twenty years the periodic orbit theory (POT) has been developed and successfully applied to low dimensional systems.
Many dynamical invariants such as natural measure, Lyapunov exponents, fractal dimensions and entropies~\cite{OttBook} can
be determined via periodic orbit (a.k.a cycle) expansions. It is an open question whether or not the POT has anything to
say for spatially extended systems. It is thus an important numerical task to find UPOs for such systems in an attempt to
answer this question. The model example of a spatiotemporally chaotic system is the Kuramoto-Sivashinsky equation (KSE),
which was first studied in the context of reaction-diffusion equations by Kuramoto and Tsuzuki~\cite{Kuramoto76}, whilst
Sivashinsky derived it independently as a model for thermal instabilities in laminar flame fronts~\cite{Sivashinsky77}. It
is one of the simplest interesting PDEs to exhibit chaos and we have chosen it as a test case for the work that follows.

The problem of finding UPOs is essentially a root finding problem, thus a popular approach is to use some variant of
Newton's method. Indeed, Zoldi and Greenside have reported the detection of $127$ distinct UPOs for the KSE~\cite{Zoldi98}.
Here they solve for the discretised system using simple shooting with a damped Newton method to update each step. Another
recent assault on the KSE by Cvitanovi\'c and Lan use variational methods~\cite{Cvitanovic03,Lan04}. Here a suitable cost
function is constructed so that its minimisation leads to the detection of a UPO. The main problem, however, is that Newton
type methods suffer from two major drawbacks: firstly, with increasing period, the basins of attraction become so small
that placing an initial seed within the basin is practically impossible, and secondly, the method has no way of
differentiating between true roots and local minima of the cost function. The latter drawback is one which increases
significantly with dimension due to the complicated topology of multi-dimensional flows.

The method of detecting UPOs by stabilising transformations~\cite{Crofts06,Lai99,Schmelcher97} aims at transforming the
system in such a way that its UPOs become stable. Unlike in the Newton-type methods, the transformations are linear and thus
do not suffer from spurious convergence. When faced with the task of finding UPOs of a discrete system
\begin{equation}\label{eqn:discrete}
x_{i+1} = f(x_i), \quad f:\mathbb{R}^n\rightarrow \mathbb{R}^n,
\end{equation}
one can look instead at the related flow
\begin{equation}\label{eqn:flow1}
\frac{dx}{ds} = g(x),
\end{equation}
where $g(x) = f^p(x) - x$. It is straightforward to see that the period$-p$ points of the map are equilibrium points for
the associated flow. With this setup we are able to stabilise all UPOs $x^*$ of Eq.~(\ref{eqn:discrete}) such that all the
eigenvalues of the Jacobian $Df^p(x^*)$ have real part smaller then one. In order to stabilise all possible UPOs we study
the following flow
\begin{equation*}
 \frac{dx}{ds} = Cg(x),
\end{equation*}
where $C\in\mathbb{R}^{n\times n}$ is a constant matrix introduced in order to stabilise UPOs with the Jacobian's that have
eigenvalues with real parts greater than one. Given a set $\{C\}$ of such matrices, we have a family of differential equations
which need to be solved in order to find all UPOs of Eq.~(\ref{eqn:discrete}). One example of such a set was proposed by
Schmelcher and Diakonos (SD)~\cite{Schmelcher97}. It is the set $\mathcal{C}_{\mathrm{SD}}$ of orthogonal matrices such that
only one entry $\{\pm1\}$ per row or column is nonzero. It has been verified that the set $\mathcal{C}_{\mathrm{SD}}$
stabilise all hyperbolic fixed points for $n\leq 2$, and numerical evidence suggests the result holds for $n>2$, but, thus
far, no proof has been presented. However, if we wish to extend the stabilising transformation approach to higher
dimensions, the set ${\mathcal C}_{\mathrm{SD}}$ cannot be applied directly, since its size increases very rapidly with
system dimension ($|\mathcal{C}_{\mathrm{SD}}|$ $= 2^nn!$).

For high-dimensional systems with relatively few unstable directions, the method of stabilising transformations can be
applied efficiently by restricting our attention to the unstable part of Eq.~(\ref{eqn:discrete}). Indeed, by
constructing transformations which only alter the stability of the flow in Eq.~(\ref{eqn:flow1}) in the unstable subspace of
$Df^p$, it is possible to reduce the number of transformations considerably. The authors have recently proposed a new set of
matrices $\mathcal{C}$ based on the properties of a small already detected set of UPOs. Here the cardinality is
$|\mathcal{C}| = 2^{n_{u}}$, where $n_u$ is the dimension of the unstable manifold at $x^*$. This is the key to extending
these ideas to higher dimensional systems, since often in practice the systems of interest are such that $n_u\ll N$. For
example, after a finite difference discretisation of the KSE with resulting system of size $N = 100$, only four of the
corresponding Lyapunov exponents are positive~\cite{Zoldi98}. Thus at each seed we would have only $|\mathcal{C}| = 16$
matrices as opposed to $|\mathcal{C}_{\mathrm{SD}}| = 2^{99}99!$ if we used the SD matrices~\footnote{Here we reduce the
flow to a Poincar\'e surface of section in order to obtain a discrete system}.

\section{Subspace decomposition}\label{sec:sub}
In what follows we take our leave from the subspace iteration methods~\cite{Lust98,Keller93}. Consider the solution of the
nonlinear system
\begin{equation}\label{eqn:nlin}
    f(x) - x = 0, \quad x\in\mathbb{R}^n, \quad
    f:\mathbb{R}^n\rightarrow\mathbb{R}^n,
\end{equation}
where $f(x)$ is assumed twice differentiable in the neighbourhood of $x^*$, an isolated root of Eq.~(\ref{eqn:nlin}). We
can approximate the solution of (\ref{eqn:nlin}) by a recursive {\em fixed point} procedure of the form
\begin{equation}\label{eqn:picard1}
    x_{i+1} = f(x_i), \quad i = 1,2,3,\dots.
\end{equation}
It is well known that the iteration (\ref{eqn:picard1}) converges locally in the neighbourhood of a solution $x^*$, as long
as all the eigenvalues of the Jacobian $Df(x^*)$ lie within the unit disc $\{z\in\mathbb{C} : |z| < 1\}$. In contrast,
(\ref{eqn:picard1}) typically diverges if $Df(x^*)$ has an eigenvalue outside the unit disc. In that case, a popular
alternative is to employ Newton's method
\begin{eqnarray}\label{eqn:newton}
    (Df(x_i) - I_n)\delta x_i &=& -(f(x_i) - x_i),\\ \nonumber
    x_{i+1} &=& x_i +\delta x_i, \quad i = 1, 2, 3, \dots.
\end{eqnarray}

The idea of subspace iterations is to exploit the fact that the divergence of the fixed point iteration (\ref{eqn:picard1})
is due to a small number of eigenvalues, $n_u$, lying outside the unit disc. By decomposing the space $\mathbb{R}^n$ into the
direct sum of the unstable subspace spanned by the eigenvectors of $Df(x^*)$
\begin{equation*}
    \mathbb{P} = \mathrm{Span}\{e_k\in\mathbb{R}^n: Df(x^*)e_k = \lambda_k e_k, |\lambda_k| > 1\}
\end{equation*}
and its orthogonal complement, $\mathbb{Q}$, a modified iterative scheme is obtained. The application of Newton's method
to the subspace $\mathbb{P}$ whilst continuing to use the relatively cheap fixed point iteration on the subspace $\mathbb{Q}$,
results in a highly efficient scheme provided $\mathrm{dim}(\mathbb{P}) \ll \mathrm{dim}(\mathbb{Q})$.

To this end, let $V_p\in\mathbb{R}^{n\times n_u}$ be a basis for the subspace $\mathbb{P}\subset\mathbb{R}^n$ spanned by
the eigenvectors of $Df(x^*)$ corresponding to those eigenvalues lying outside the unit disc, and
$V_q\in\mathbb{R}^{n\times n_s}$ a basis for $\mathbb{Q}$, where $n_u+n_s=n$. Then, we can define orthogonal projectors $P$
and $Q$ onto the respective subspaces, $\mathbb{P}, \mathbb{Q}$, as follows
\begin{eqnarray}\nonumber
  P &=& V_pV_p^{\mathsf{T}},\\
  \label{eqn:orthog}
  Q &=& V_qV_q^{\mathsf{T}} = I_n - P.
\end{eqnarray}
Note that any $x\in\mathbb{R}^n$ admits the following unique decomposition
\begin{equation}\label{eqn:xdec}
    x = V_p\bar{p} + V_q\bar{q} = p + q, \quad p := V_p\bar{p} = Px,
    \quad q := V_q\bar{q} = Qx,
\end{equation}
with $\bar{p}\in\mathbb{R}^{n_u}$ and $\bar{q}\in\mathbb{R}^{n_s}$. Substituting (\ref{eqn:xdec}) in Eq.~(\ref{eqn:newton})
and multiplying the result by $[V_q, V_p]^{\mathsf{T}}$ on the left, one obtains
\begin{equation}\label{eqn:newtsub}
\left[
  \begin{array}{cc}
    V_q^{\mathsf{T}}DfV_q - I_{n_s} & 0 \\
    V_p^{\mathsf{T}}DfV_q & V_p^{\mathsf{T}}DfV_p - I_{n_u} \\
  \end{array}
\right] \left[
  \begin{array}{c}
    \Delta\bar{q} \\
    \Delta\bar{p} \\
  \end{array}
\right] = -\left[
  \begin{array}{c}
    V_q^{\mathsf{T}}f - \bar{q} \\
    V_p^{\mathsf{T}}f - \bar{p}\\
  \end{array}
\right].
\end{equation}
Here we have used the fact that $V_p^{\mathsf{T}}V_q = 0_{n_u\times n_s}$, $V_q^{\mathsf{T}}V_p = 0_{n_s\times n_u}$, and
$V_q^{\mathsf{T}}DfV_p = 0_{n_s\times n_u}$ the latter holding due to the invariance of $Df$ on the subspace $\mathbb{P}$.
Now, the first $n_s$ equations in (\ref{eqn:newtsub}) may be solved using the following fixed point iteration scheme
\begin{eqnarray}\label{eqn:picard2}\nonumber
  \Delta\bar{q}^{[0]} &=& 0, \\
  \Delta\bar{q}^{[i]} &=& V_q^{\mathsf{T}}DfV_q\Delta\bar{q}^{[i-1]} + V_q^{\mathsf{T}}f - \bar{q},
  \\\nonumber
  \Delta\bar{q} &=& \Delta\bar{q}^{[l]} =
  \sum_{i=0}^{l-1}(V_q^{\mathsf{T}}DfV_q)^i(V_q^{\mathsf{T}}f - \bar{q}),
\end{eqnarray}
where $l$ denotes the number of fixed point iterations taken per NR step. Since $r_{\sigma}[V_q^{\mathsf{T}}DfV_q] < 1$ by
construction, the iteration (\ref{eqn:picard2}) will be locally convergent on $\mathbb{Q}$ in some neigbourhood of
$\Delta\bar{q}$ -- here $r_{\sigma}[\cdot]$ denotes the spectral radius. In order to determine $\Delta\bar{p}$ one solves
\begin{equation*}
    (V_p^{\mathsf{T}}DfV_p - I_{n_u})\Delta\bar{p} =
    -V_p^{\mathsf{T}}f + \bar{p} - V_p^{\mathsf{T}}DfV_q\Delta\bar{q}.
\end{equation*}
Note that in practice only one iteration of (\ref{eqn:picard2}) is performed~\cite{Lust98}, i.e. $l = 1$, this leads to
the following simplified system to solve for the correction $[\Delta\bar{q}, \Delta\bar{p}]^{\mathsf{T}}$
\begin{equation*}
\left[\begin{array}{cc}
    -I_{n_s} & 0 \\
    V_p^{\mathsf{T}}DfV_q & V_p^{\mathsf{T}}DfV_p - I_{n_u} \\
  \end{array}
\right] \left[\begin{array}{c}
    \Delta\bar{q} \\
    \Delta\bar{p} \\
  \end{array}
\right] = - \left[\begin{array}{c}
    V_q^{\mathsf{T}}f - \bar{q} \\
    V_p^{\mathsf{T}}f - \bar{p}\\
  \end{array}
\right].
\end{equation*}

Key to the success of the above algorithm is the accurate approximation of the eigenspace corresponding to the unstable
modes. In order to construct the projectors $P$, $Q$, the Schur decomposition is used. However, primary concern of the
work in~\cite{Lust98,Keller93} is the continuation of branches of periodic orbits, where it is assumed that a reasonable approximation to a UPO is known. Since we have no knowledge {\em apriori} of the orbits whereabouts we shall need to accommodate this into our extension of the method to detecting UPOs.

\subsection{Stabilising transformations}\label{sec:stabtrans}
As discussed in the introduction, an alternative approach is supplied by the method of STs, where in order to detect equilibrium solutions of Eq.~(\ref{eqn:nlin}) we introduce the associated flow
\begin{equation}\label{eqn:flow2}
    \frac{dx}{ds} = Cg(x).
\end{equation}
Here $g = f^p(x)-x$ and $C\in\mathbb{R}^{n\times n}$ is a constant
matrix.

Now, substituting (\ref{eqn:xdec}) in Eq.~(\ref{eqn:flow2}) and multiplying the result by $[V_q, V_p]^{\mathsf{T}}$ on the left, one obtains
\begin{eqnarray}\label{eqn:1}
  \frac{d\bar{q}}{ds} &=&   V_q^{\mathsf{T}}g, \\\label{eqn:2}
  \frac{d\bar{p}}{ds} &=& V_p^{\mathsf{T}}g.
\end{eqnarray}
Thus we have replaced the original Eq.~(\ref{eqn:flow2}) by a pair of coupled equations, Eq.~(\ref{eqn:1}) of dimension $n_s$ and Eq.~(\ref{eqn:2}) of dimension $n_u$. Since
\begin{eqnarray*}
  \frac{\partial}{\partial\bar{q}}(V_q^{\mathsf{T}}g) &=& V_qDg\frac{\partial x}{\partial\bar{q}}, \\
            &=& V_q^{\mathsf{T}}DgV_q,
\end{eqnarray*}
and $r_{\sigma}[V_q^{\mathsf{T}}DgV_q] < 0$ by construction, it follows that in order to detect all UPOs of Eq.~(\ref{eqn:nlin}), it is sufficient to solve
\begin{eqnarray}\label{eqn:3}
  \frac{d\bar{q}}{ds} &=& V_q^{\mathsf{T}}g, \\\label{eqn:4}
  \frac{d\bar{p}}{ds} &=& \tilde{C}V_p^{\mathsf{T}}g,
\end{eqnarray}
where $\tilde{C}\in\mathbb{R}^{n_u\times n_u}$ is a constant matrix.
\begin{figure}[t]
\begin{center}
\subfigure[Schur decomposition]{
\includegraphics[scale=0.75]{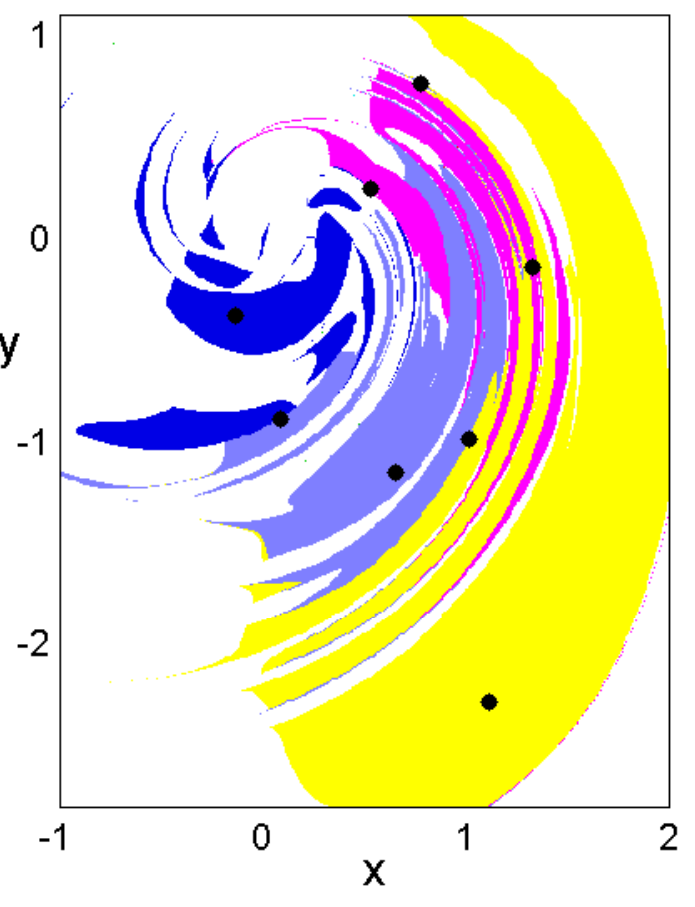}}
\hspace{0.5cm} \subfigure[Singular value decomposition]{
\includegraphics[scale=0.75]{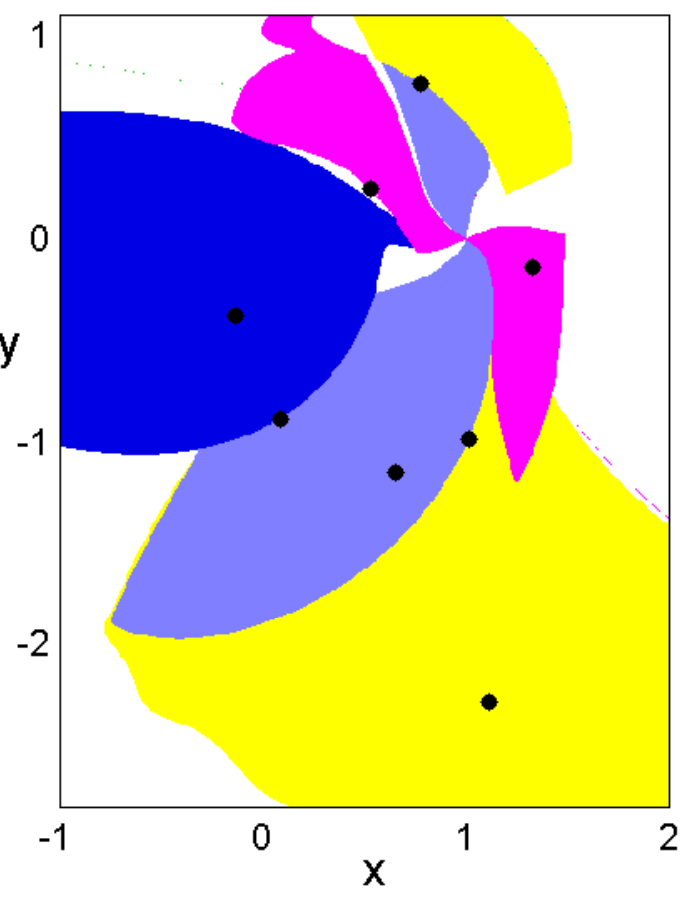}}
\caption{Basins of attraction for the period$-3$ orbits of the Ikeda
map with parameter values $a = 1.0$, $b = 0.9$, $k = 0.4$ and $\eta
= 6.0$. Here we have chosen $\tilde{C} = -1$ since in this example
the unstable subspace is one-dimensional.} \label{fig:basin}
\end{center}
\end{figure}

In~\cite{Lust98,Keller93} the Schur decomposition is used in order to construct the projectors $P$ and $Q$. This is fine for continuation problems since one may assume from the offset that they possess an initial condition $x_0$ sufficiently close to a UPO such that the Schur decomposition of $Df(x_0)$ gives a good approximation to the eigenspace of $Df(x^*)$. However, it is well known that the eigenvectors of the perturbed Jacobian $Df(x^*+\delta x)$ behave
erratically as we increase $\delta x$. In order to enlarge the basins of attraction for the UPOs we propose that singular value decomposition (SVD) be used instead. That is we choose an initial condition $x_0$ and construct the SVD of its pre-image, i.e. $Df(f^{-1}(x_0)) = USW^{\mathsf{T}}$ (or in the continuous case $D\phi^T(\phi^{-T}(x_0)) = USW^{\mathsf{T}}$ for some time $T$), the columns of $U$ give the stretching directions of the map at $x_0$,
whilst the singular values determine whether the directions are expanding or contracting. It is these directions which we use to construct the projectors $P$ and $Q$. Due to the robustness of the SVD we expect the respective basins of attraction to increase.

It is not necessary in practice to decompose Eq.~(\ref{eqn:flow2}) in order to apply the new ST. Rather we can express $C$ in terms of $\tilde{C}$ and $V_p$. To see this we add $V_q$ times Eq.~(\ref{eqn:3}) to $V_p$ times Eq.~(\ref{eqn:4}) to get
\begin{eqnarray*}
\frac{dx}{ds} &=& V_qV_q^{\mathsf{T}}g(x) + V_p\tilde{C}V_p^{\mathsf{T}}g(x),\\
              &=& [I_n + V_p(\tilde{C} - I_{n_u})V_p^{\mathsf{T}}]g(x),
\end{eqnarray*}
where the second line follows from Eq.~(\ref{eqn:orthog}). From this we see that the following choice of $C$ is equivalent to the preceding decomposition
\begin{equation}\label{eqn:Cmat}
C = I_n + V_p(\tilde{C} - I_{n_u})V_p^{\mathsf{T}}.
\end{equation}
Thus in practice we compute $V_p$ and $\tilde{C}$ at the seed $x_0$ in order to construct $C$ and then proceed to solve
Eq.~(\ref{eqn:flow2}).

\begin{figure}[t]
\begin{center}
\includegraphics[scale=0.8]{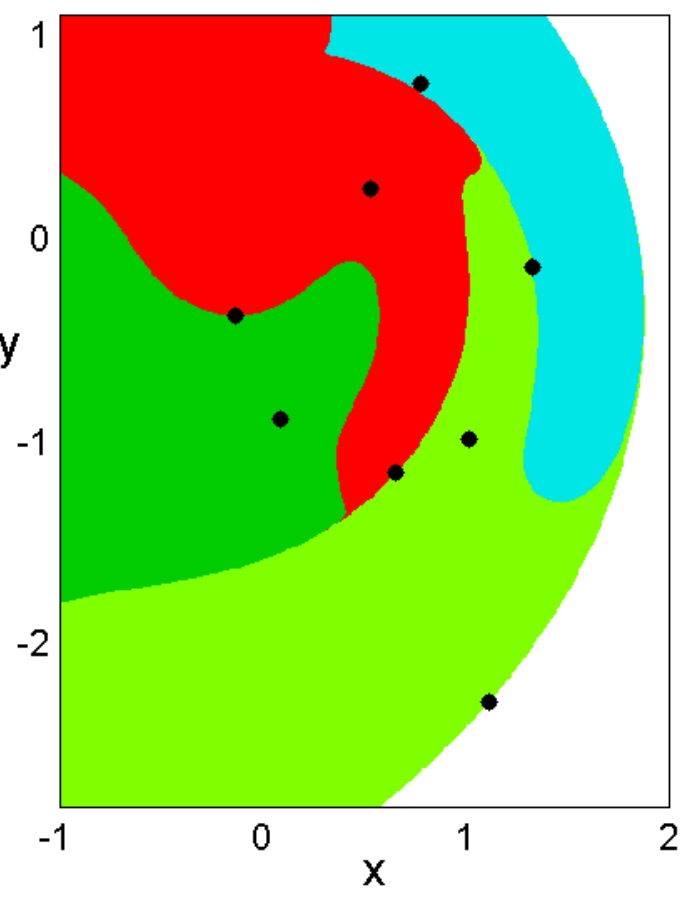}
\caption{The basins of attraction for the Ikeda map for the choice
of $\tilde{C} = 1$. Fixed points of $f^3$ with negative unstable
eigenvalues are stable stationary solutions of the associated flow,
while those with positive eigenvalues are saddles located at the
basin boundaries.} \label{fig:id}
\end{center}
\end{figure}

The advantage of using the SVD rather than the Schur decomposition can be illustrated by the following example. Consider the Ikeda map~\cite{Ikeda79}:
\begin{equation}
f(\mathbf{x}) := \left[\!\!\begin{array}{c} x_{i+1}\\
y_{i+1}\end{array}
\!\!\right] = \left[\!\!\begin{array}{c} a + b(x_i\cos{(\phi_i)} - y_i\sin{(\phi_i)})\\
 b(x_i\sin{(\phi_i)} + y_i\cos{(\phi_i)})\end{array}\!\!\right],\label{eqn:ikeda}
\end{equation}
where $\phi_i = k-\eta/(1+x_i^2+y_i^2)$ and the parameters are chosen such that the map has a chaotic attractor: $a=1.0$, $b=0.9$, $k=0.4$ and $\eta = 6.0$.  For this choice of parameters the Ikeda map possesses eight period$-3$ orbit points (two period-3 orbits and two fixed points, one of which is on the attractor basin boundary). In our experiments we have covered the attractor for Eq.~(\ref{eqn:ikeda}) with a grid of initial seeds and solved
the associated flow for $p=3$, i.e., $g(x) = f^3(x) - x$. This is done twice, firstly in the case where the projections $P$ and $Q$ are constructed via the Schur decomposition and secondly when they are constructed through the SVD. Since all UPOs of the Ikeda map are of saddle type, the unstable subspace is one-dimensional and we need only two transformations: $\tilde{C} = 1$ and $\tilde{C} = -1$. Figure \ref{fig:basin} shows the respective basins of attraction
for the two experiments with $\tilde{C} = -1$. It can be clearly seen that the use of SVD corresponds to a significant increase in basin size compared to the Schur decomposition. Note that with $\tilde{C} = -1$ we stabilise four out of eight fixed points of $f^3$. The other four are stabilised with $\tilde{C} = 1$. The corresponding basins of attraction are shown in Figure \ref{fig:id}. Note that for this choice of ST the decision of which basis vectors to choose is not important, since Eq.~(\ref{eqn:Cmat}) yields $C = I$, so that the associated flow is independent of the
selected basis.

\section{Implementation} \label{sec:imp}
A typical approach in the determination of UPOs for flows is via a Poincar\'e surface of section (PSS). By ``clever'' placement of an $(n-1)$-dimensional manifold in the phase space, the problem is reduced to a discrete map defined via intersections with the manifold. However, a correct choice of PSS is a challenging problem in itself. Due to the complex topology of a high-dimensional phase space, the successful detection of UPOs will be highly dependent upon the choice of surface. When the choice of a suitable PSS is not obvious {\em a priori}, we found it preferable to work with the full
flow, adding an auxiliary equation to determine the integration time $T$.

Let $x\mapsto\phi^t(x)$ denote the flow map of Eq.~(\ref{eqn:ode}). Then we define the associated flow as follows
\begin{equation}\label{eqn:assflow}
    \frac{dx}{ds} = C(\phi^T(x) - x).
\end{equation}
The additional equation for $T$ is constructed such that $T$ is always changing in the direction that decreases $|\phi^T(x) - x|$, i.e.
\begin{equation*}
    \frac{dT}{ds} \varpropto -\frac{\partial |g|^2}{\partial T},
\end{equation*}
or, more precisely
\begin{equation}\label{eqn:Teqn}
    \frac{dT}{ds} = -\alpha
    v(\phi^T(x))\cdot(\phi^T(x)-x).
\end{equation}
Here $\alpha > 0$ is a constant which controls the relative speed of convergence of Eq.~(\ref{eqn:Teqn}). This leads to the following augmented flow which must be solved to detect UPOs of (\ref{eqn:ode}):
\begin{equation}\label{eqn:augflow}
\frac{d}{ds}
    \left[
      \begin{array}{c}
        x \\
        T \\
      \end{array}
    \right] =
    \left[
      \begin{array}{c}
        C(\phi^T(x) - x) \\
        -\alpha v(\phi^T(x))\cdot(\phi^T(x) - x) \\
      \end{array}
    \right].
\end{equation}
Note that the augmented flow (\ref{eqn:augflow}) is derived by integrating a nonlinear PDE for some time $T$ and will become increasingly stiff for larger $T$.

Several approaches have been proposed for the solution of stiff systems of ODEs; see for example, the review by Shampine and Gear \cite{Gear79}. Of all these techniques the general-purpose codes contained within the ODEPACK software package \cite{ODEPACK} are regarded as some of the best available routines for the solution of such systems.  Thus, in our numerical experiments we use the stiff solver {\tt dlsodar} from the ODEPACK toolbox to integrate (\ref{eqn:augflow}). {\tt dlsodar} is a variable step-size solver which automatically changes between stiff and nonstiff modes. In particular, as we approach a steady state of Eq.~(\ref{eqn:augflow}) {\tt dlsodar} will take increasingly larger time-steps, leading to super linear convergence in the neighbourhood of the solution.

To use the solver {\tt dlsodar}, we must provide a routine that returns the value of the vector field (\ref{eqn:augflow}) evaluated at a given point $(x, T)^{\mathsf{T}}$. Here we need the solution of Eq.~(\ref{eqn:ode}) which is obtained by applying a suitable numerical integration scheme; see the next section for further details. The ODEPACK software package makes use of the Jacobian matrix of the system being integrated and provides the
option of computing the Jacobian via finite differences or via a user supplied routine. Note that in the case that the flow is expected to be stiff much of the time, it is recommended that a routine for the Jacobian is supplied and we do this. The derivative of (\ref{eqn:augflow}) with respect to $(x, T)^{\mathsf{T}}$ is given by
\begin{equation*}
 \left[ \begin{array}{cc}
  C(J_T - I_n) & Cv_T \\
  -\alpha( v_T^{\mathsf{T}}(J_T - I_n) + g^{\mathsf{T}}DvJ_T) & -\alpha(g^{\mathsf{T}}Dvv_T
  + v_T^{\mathsf{T}}v_T) \\
 \end{array}\right],
\end{equation*}
where $J_T = \partial\phi^T(x)/\partial x $, $v_T = v(\phi^T(x))$, $Dv = dv/dx$ and $g = \phi^T(x) - x$ as usual.

Quite often one might wish to terminate simulation before the usual stopping criteria of, for example, a maximum number of steps taken or certain tolerances having been reached. A nice feature of the {\tt dlsodar} algorithm is that it allows the optional user supplied routine to do just this. To be more exact, it determines the roots of any of a set of user supplied functions
\begin{equation*}
h_i = h_i(t,x_1,\dots,x_n),\quad i = 1,\dots,m,
\end{equation*}
and returns the solution of (\ref{eqn:augflow}) at the root if it occurs prior to the normal stopping criteria.

Note that, to increase the efficiency of the algorithm we wish to avoid the following two instances: firstly, due to the local nature of the STs, we should stop the search if we wander too far from the initial condition, and secondly, since our search is governed by the dynamics of Eq.~(\ref{eqn:assflow}) and not by those of (\ref{eqn:ode}), we might move off the attractor after some time period so the convergence to a UPO becomes highly unlikely. In our numerical experiments we supply the following function
\begin{equation*}
    h_1  = a - ||g||,
\end{equation*}
where $a\in\mathbb{R}$ is a constant and $||\cdot||$ denotes the $L_2$ norm. In practice, we have found that there exists a threshold value of $a$, such that convergence is highly unlikely once the norm of $g$ surpasses it. Note that we also restrict the maximum number of allowed integration steps since the convergence becomes less likely if the associated flow is integrated for a long time.

We must also provide two tolerances, rtol and atol, which control the local error of the ODE solver. In particular, the estimated local error in $X = (x, T)^{\mathsf{T}}$ will be controlled so as to be less than
\begin{equation*}
\mathrm{rtol}\cdot||X||_{\infty} + \mathrm{atol}.
\end{equation*}
Thus the local error test passes if, in each component, either the absolute error is less than atol or the relative error is less than rtol. The accuracy with which we would like to solve the flow (\ref{eqn:augflow}) is determined by the stability properties of (\ref{eqn:ode}). To understand this, we note that in evaluating the RHS of (\ref{eqn:augflow}) it is the solution of Eq.~(\ref{eqn:ode}) at time $T$, i.e. $\phi^T(x)$, which is critical
for error considerations. Suppose that our initial point lies within $\delta x$ of a true trajectory $x$. Then $\phi^T(x+\delta x)$ lies approximately within $e^{\lambda T}\delta x$ of the true trajectory, $\phi^T(x)$, where $\lambda$ is the largest Lyapunov exponent of the system. Since $\lambda$ is positive for chaotic systems, the error
grows exponentially with the period and we should take this into account when setting the tolerances rtol and atol. This leads us to the following settings for the tolerances
\begin{equation*}
    \mathrm{rtol} = \mathrm{atol} = 10^{-5}/e^{\lambda T_0},
\end{equation*}
where $T_0$ is the initial period and $\lambda$ is the largest Lyapunov exponent of the flow $v$. We have computed the Lyapunov exponent using the algorithm due to Benettin {\em et al} \cite{Benettin80}.

\subsection{Kuramoto-Sivashinsky equation}\label{sec:kse}
We have chosen the Kuramoto-Sivashinsky equation (KSE) for our numerical experiments. It is the simplest example of spatiotemporal chaos and has been studied in a similar context in~\cite{Christiansen97,Lan04,Zoldi98}, where the detection of many UPOs has been reported. We work with the KSE in the form
\begin{equation}\label{eqn:kse}
    u_t = -\frac{1}{2}(u^2)_x - u_{xx} - u_{xxxx},
\end{equation}
where $x\in [0, L]$ is the spatial coordinate, $t\in\mathbb{R}^{+}$ is the time and the subscripts $x$, $t$ denote differentiation with respect to space and time.  For $L < 2\pi$, $u(x,t) = 0$ is the
global attractor for the system and the resulting long time dynamics are trivial. However, for increasing $L$ the system undergoes a sequence of bifurcations leading to complicated dynamics; see for example~\cite{Kevrekidis90}.

Our setup will be close to that found in~\cite{Lan04}. In what follows we assume periodic boundary conditions: $u(x,t) = u(x+L,t)$, and restrict our search to the subspace of antisymmetric solutions, i.e. $u(x,t) = -u(L-x,t)$. Due to the periodicity of the solution, we can solve Eq.~(\ref{eqn:kse}) using the pseudo-spectral method~\cite{HairerBook,TrefethenBook}. Representing the function $u(x,t)$ in terms of its Fourier modes:
\begin{equation*}
  \quad u(x,t) := {\cal F}^{-1}[\hat{u}] = \sum_{k\in{\mathbb Z}}
\hat{u}_k e^{-ikqx},
\end{equation*}
where
\begin{equation*}
\hat{u}:=(\dots,\hat{u}_{-1},\hat{u}_0,\hat{u}_1,\dots)^{\mathsf{T}}\,,\qquad
     \hat{u}_k := {\cal F}[u]_k = \frac{1}{L}\int_0^L u(x,t)
e^{ikqx}dx,
\end{equation*}
we arrive at the following system of ODEs
\begin{equation*}
  \frac{d\hat{u}_k}{dt} =[(kq)^2-(kq)^4]\hat{u}_k -
  \frac{ikq}{2}{\cal F}[({\cal F}^{-1}[\hat{u}])^2]_k\,.
\end{equation*}
Here $q = 2\pi/L$ is the basic wave number. Since $u$ is real, the Fourier modes are related by $\hat{u}_{-k} = \hat{u}^\ast_k$. Furthermore, since we restrict our search to the subspace of odd solutions, the Fourier modes are pure imaginary, i.e. $\mathfrak{Re}(\hat{u}_{k}) = 0$.

The above system is truncated as follows: the Fourier transform ${\cal F}$ is replaced by its discrete equivalent
\begin{equation*}
  a_k := {\cal F}_N[u]_k = \sum_{j = 0}^{N-1} u(x_j)
  e^{ikqx_j}\,,\qquad u(x_j) := {\cal F}_N^{-1}[a]_j
  = \frac{1}{N}\sum_{k = 0}^{N-1} a_j e^{-ikqx_j}\,,
\end{equation*}
where $x_j = L/N$ and $a_{N-k} = a^\ast_k$. Since $a_0 = 0$ due to Galilean invariance and setting $a_{N/2} = 0$ (assuming $N$ is even), the number of independent variables in the truncated system is $n = N/2-1$.  The truncated system looks as follows:
\begin{equation}\label{eqn:finite}
  \dot{a}_k =[(kq)^2-(kq)^4]a_k -
  \frac{ikq}{2}{\cal F}_N[({\cal F}_N^{-1}[a])^2]_k\,,
\end{equation}
with $k = 1,\ldots,n$, although in the Fourier transform we need to use $a_k$ over the full range of $k$ values from 0 to $N-1$.

The discrete Fourier transform ${\cal F}_N$ can be computed using fast Fourier transform (FFT). In Fortran and C, the routine {\tt REALFT} from Numerical Recipes~\cite{NumericalRecipes} can be used. In Matlab, it is more convenient to use complex variables for $a_k$. Note that Matlab function {\tt fft} is, in fact, the inverse Fourier transform.

To derive the equation for the matrix of variations, we use the fact that ${\cal F}_N$ is a linear operator to obtain
\begin{equation}\label{eqn:varform}
  \frac{\partial \dot{a}_k}{\partial a_j} =
  [(kq)^2-(kq)^4]\delta_{kj} +
  kq{\cal F}_N[{\cal F}_N^{-1}[a]\otimes{\cal
  F}_N^{-1}[\delta_{kj}]]\,,\quad j = 1,\ldots,N-2,
\end{equation}
where $\otimes$ indicates componentwise product, and the inverse Fourier transform is applied separately to each column of $\delta_{kj}$. Here, $\delta_{kj}$ is not a standard Kronecker delta, but the $N\times n$ matrix:
\begin{equation*}
  \delta_{kj} = \left(
  \begin{array}{ccc}
  0 &  0 &\cdots\\
  1 &  0 &\cdots\\
  0 &  1 &\cdots\\
\multicolumn{3}{c}\dotfill \\
  0 & 0 &\cdots\\
\multicolumn{3}{c}\dotfill \\
  0  & -1 &\cdots\\
  -1 & 0 &\cdots\\
  \end{array}  \right),
\end{equation*}
with index $k$ running from 0 to $N-1$.

In practice, the number of degrees of freedom $n$ should be sufficiently large so that no modes important to the dynamics are truncated, whilst on the other hand, an increase in $n$ corresponds to an increase in computation. To determine the order of the truncation in our numerical experiments, we initially chose $n$ to be large and integrated a random initial seed onto the attractor. By monitoring the magnitude of the harmonics an integer $k$ was determined such that $a_j<10^{-5}$ for $j>k$. The value of $n$ was then chosen to be the smallest integer such that: (i) $n\geq k$, and (ii) $N = 2n+2$ was an integer power of two. The second condition ensures that the FFT is applied to vectors of size which is a power of two resulting in optimal performance.

Note that in the numerical results to follow we work entirely in Fourier space. We use an exponential time differencing method (ETDRK4) due to Kassam and Trefethen~\cite{Kassam05} in order to solve (\ref{eqn:finite}) and (\ref{eqn:varform}). Note that the method uses a fixed step-size ($h = 0.25$ in our calculations) thus it is necessary to use an interpolation scheme in order to integrate up to arbitrary times. In our work we have implemented cubic interpolation~\cite{NumericalRecipes}. More precisely, to integrate up to time $t\in[t_i, t_i+h]$, where the $t_i$ are integer multiples of the step-size $h$. We construct the unique third order polynomial passing through the two points $a(t_i)$ and $a(t_i+h)$, with derivatives $a'(t_i)$ and $a'(t_i+h)$ at the respective points. In this way we obtain the following cubic model:
\begin{align*}
    p(s) = [&\,2a(t_i)+a'(t_i)+a'(t_i+h)-2a(t_i+h)]s^3 +
           [3a(t_i+h)-3a(t_i)\notag\\&-2a'(t_i)-a'(t_i+h)]s^2 +
           a'(t_i)s + a(t_i),
\end{align*}
where the parameter $s = (t-t_i)/h\in[0, 1]$.

\subsection{Numerical Results}\label{sec:num}

We now present the results of our numerical experiments. We have compared the performance of our method against the Newton-Armijo (NA) version of the damped Newton algorithm~\cite{KelleyBook}, as well as the nonlinear least squares solver {\tt lmder} from the MINPACK software package \cite{Minpack}. Note that {\tt lmder} is an implementation of the Levenburg-Marquardt algorithm~\cite{Marquardt63}. Both methods have been successfully applied to spatially extended systems in order to detect periodic orbits in the past, in particular, in \cite{Zoldi98} the NA algorithm was able to detect many distinct UPOs of the KSE, whilst more recently, {\tt lmder} has been used to determine many UPOs of the closely related complex Ginzburg-Landau equation~\cite{Lopez05}.

Both methods require as input a function whose zeros are to be determined. In the case of NA an auxiliary equation must be added due to the time invariance of Eq.~(\ref{eqn:ode}). For {\tt lmder}, it is not necessary to augment the system since the algorithm is able to solve under-determined systems of equations, however, we run two separate experiments: (i) we solve the unconstrained system, and (ii) we supply an auxiliary equation as in the case of the Newton algorithm. In this way we are able to infer the effect of the PSS on the performance of the search. One important note concerning
{\tt lmder}, is that, even though it is capable of solving under-determined systems of equations, it still requires the number of equations to equal the number of unknowns in the user supplied function, in practice however, we may simply set any additional equations identically equal to zero.

In order to determine UPOs via the aforementioned methods, we introduce the following augmented system
\begin{equation}\label{eqn:augsys}
F(x,T) =  \left[
      \begin{array}{c}
        \phi^T(x) - x \\
        v(x_0)\cdot(\phi^T(x) - x_0)\\
      \end{array}
    \right],
\end{equation}
where the additional equation defines a Poincar\'{e} surface of section normal to the initial field vector. Also, in order to optimise efficiency we use the analytic Jacobian in all our experiments rather than a numerical approximation
\begin{equation}\label{eqn:augsysdf}
 DF(x,T) = \left[
  \begin{array}{cc}
   J_T - I_n & v_T\\
   v_0       & 0
   \end{array}
 \right].
\end{equation}
Note that, in the case of the unconstrained system, the equations are as above, except that we set the last entry in (\ref{eqn:augsys}) and the final row in (\ref{eqn:augsysdf}) identically equal to zero. In addition to this we must supply the two routines with certain tolerances in order to control errors. Let us denote the relative error desired in the approximate solution by xtol and the relative error desired in the sum of squares by ftol. In the computations performed in the next section we set the tolerances to the recommended values of
\begin{equation*}
\mathrm{ftol} = 10^{-8}, \quad \mathrm{xtol} = 10^{-8}.
\end{equation*}
Further, {\tt lmder} requires a third tolerance, gtol, which measures the orthogonality desired between the vector function $F$ and the columns of the Jacobian, $DF$; we also set this to its recommended value gtol$=0$. Finally, we specify a maximum number of function evaluations allowed during each run of the {\tt lmder} and NA algorithms in order to increase efficiency.

For our method we use the set of matrices proposed in \S\ref{sec:stabtrans} with
\begin{equation*}
\tilde{C} = \mathcal{C}_{\mathrm{SD}},
\end{equation*}
since within the low-dimensional unstable subspace it is possible to apply the full set of Schmelcher-Diakonos (SD) matrices. The UPOs determined from our search will then be used as seeds to determine new cycles. Here we proceed in analogous fashion to \cite{Crofts06} by constructing STs from the monodromy matrix, $D\phi^{T^*}(a^*)$, of the cycle
$(a^*, T^*)^{\mathsf{T}}$. We then solve the augmented flow (\ref{eqn:augflow}) from the new initial condition $(a^*, \tilde{T})^{\mathsf{T}}$, where the time $\tilde{T}$ is chosen such that
\begin{equation*}
a^*(0) \approx a^*(\tilde{T}) \quad \mathrm{and} \quad \tilde{T}~(\mathrm{mod}T^*) \neq 0.
\end{equation*}
Note that any given cycle may exhibit many close returns, particularly longer cycles, thus in general a periodic orbit may produce many new initial seeds. This is an especially useful feature, since we do not have to recompute the corresponding STs.

Initially, to determine that a newly detected cycle, $(x^*, T^*)$, was different from those already found, we first checked whether the periods differed, that is, we determined whether $|T^* - T'| > T_{\mathrm{tol}}$ for all previously detected orbits. If two orbits where found to have the same period, we then calculated the distance between the first components of $x^*$, and all other detected orbits $y^*$. However, in practice we found, that if two orbits have the same period, then either they are the same or they are related via symmetry, recall that if $u(x,t)$ is a solution then so is $-u(L-x,t)$. Thus, in order to avoid the convergence of UPOs that are trivially related by symmetry we will consider two orbits as being equal if their periods differ by less than the tolerance $T_{\mathrm{tol}}$. Of course, this criterion can, in theory, lead us to discard cycles incorrectly, however, this is highly unlikely in practice.

\subsubsection{Comparison of the numerical methods}
The search for UPOs is conducted within a rectangular region containing the chaotic invariant set. Initial seeds are obtained by integrating a random point within the region for some transient time $\tau$. Once on the attractor, the search for close returns within chaotic dynamics is implemented. That is, we integrate the system from the initial point on the attractor until $a(t_0) \approx a(t_1)$ for some times $t_0 < t_1$, and use the close return, $(a(t_0), T_0)$, where $T_0 = t_1 - t_0$, as our initial guess to a time-periodic solution. In order to build the STs we solve the variational equations for each seed starting from the random initial point, $a(0)$, for time $\tau + t_0$. In order to construct the matrix $V_p$, we apply the SVD to the matrix
\begin{equation}\label{eqn:dfsvd}
    D\phi^{\tau+t_0}(a(0))= USW^{\mathsf{T}},
\end{equation}
the corresponding ST is given by
\begin{equation*}\label{eqn:Cmatt}
    C = I_n + V_p(\tilde{C} - I_{n_u})V_p^{\mathsf{T}},
\end{equation*}
where $V_p = U_{jk}$, $j=1,\dots,n$, $k = 1,\dots,n_u$, i.e. the first $n_u$ columns of $U$ in (\ref{eqn:dfsvd}). Here $n_u$ is the number of expanding directions which is determined by the number of singular values with modulus greater than one.

\begin{table}
 \caption{The number of distinct periodic solutions for the
 Kuramoto-Sivashinsky equation detected by the method of STs.
 Here $L = 38.5$ and $\alpha = 0.25$.}
    \begin{center}
        \begin{tabular}{|c|c||c|c|c|c|c|}
            \hline
            Period & C & $N_{\mathrm{po}}$ & $N_{\mathrm{hit}}$ & $N_{\mathrm{fev}}$
            & $N_{\mathrm{jev}}$ & Work\\
            \hline
            \multirow{3}{*}{$10 - 100$} & $C_0$ & $28$ & $498$ & $252$ & $10$ & $412$ \\
            &  $C_1$ & $16$ & $296$ & $684$ & $32$ & $1196$
            \\ \cline{3-7}
            &  $\{C_i\}$ & $44$ & $794$ & $936$ & $42$ & $1608$ \\ \hline
            \hline
            \multirow{3}{*}{$100 - 250$} & $C_0$ & $235$ & $395$ & $903$ & $78$ & $2151$ \\
            &  $C_1$ & $64$ & $256$ & $1294$ & $112$ & $3086$
            \\ \cline{3-7}
            &  $\{C_i\}$ & $299$ & $651$ & $2197$ & $190$ & $5237$ \\ \hline
        \end{tabular}
    \end{center}\label{table1}
\end{table}

\begin{table}[t]
 \caption{The number of distinct periodic solutions for the Kuramoto-Sivashinsky
 equation detected by the Levernberg-Marquardt algorithm {\tt lmder} with $L = 38.5$.}
    \begin{center}
        \begin{tabular}{|c|c||c|c|c|c|c|}
            \hline
            Method & Period & $N_{\mathrm{po}}$ & $N_{\mathrm{hit}}$ & $N_{\mathrm{fev}}$
            & $N_{\mathrm{jev}}$ & Work\\
            \hline
            \multirow{2}{*}{Constrained} & $10-100$ & $41$ & $475$ & $12$ & $9$ & $156$ \\
            &  $100-250$ & $232$ & $358$ & $113$ & $97$ & $1665$ \\ \hline
            \hline
            \multirow{2}{*}{Unconstrained} & $10-100$ & $42$ & $491$ & $11$ & $9$ & $155$ \\
            &  $100-250$ & $233$ & $363$ & $109$ & $95$ & $1629$ \\ \hline
        \end{tabular}
    \end{center}\label{table2}
\end{table}

\begin{table}[t]
 \caption{The number of distinct periodic solutions for the Kuramoto-Sivashinsky
 equation detected by the Newton-Armijo algorithm with $L = 38.5$.}
    \begin{center}
         \begin{tabular}{|c||c|c|c|c|c|}
            \hline
            Period & $N_{\mathrm{po}}$ & $N_{\mathrm{hit}}$ & $N_{\mathrm{fev}}$ & $N_{\mathrm{jev}}$ &
            Work\\\hline
            $10 - 100$  & $43$ & $475$ & $29$ & $7$ & $141$ \\
            $100 - 250$ & $213$ & $334$ & $309$ & $41$ & $965$ \\
            \hline
        \end{tabular}
    \end{center}\label{table3}
\end{table}

We examine two different system sizes: $L = 38.5$, for which the detected UPOs typically have one positive Lyapunov exponent, and $L = 51.4$, for which the UPOs have either one or two positive Lyapunov exponents. The corresponding systems sizes are $n = 15$ and $n =31$ respectively. Our experiments where conducted over two separate ranges. We began by looking for shorter cycles with period $T\in[10, 100]$, the lower bound here was determined {\em a posteriori} so as to be smaller than the shortest detected cycle. We then searched for longer cycles, $T\in[100, 250]$ to be more precise, where the maximum of $T=250$ was chosen in order to reduce the computational effort.

In our calculations we set the positive constant $\alpha=0.25$ in Eq.~(\ref{eqn:augflow}). Using the solver {\tt dlsodar} we integrated $500$ random seeds over both ranges for time $s = 150$, the seeds where chosen such that $|\phi^{T_0}(a(0))-a(0)| < 1.0$. If the flow did not converge within $1000$ integration steps, we found it more efficient to terminate the solver and to re-start with a different ST or a new seed. As mentioned in the previous section we choose a constant $a = 50$ experimentally so that integration is terminated if the norm of $g$ grows to large, i.e. $|g(x)| = |\phi^T(x) - x| > a$. Typically on convergence of the associated flow the UPO is determined with accuracy of about $10^{-7}$ at which point we implement two or three iterations of the Newton-Armijo rule to Eq.~(\ref{eqn:augsys}) in order to allow convergence to a UPO to within roundoff error.

\begin{figure}[t]
\begin{center}
\includegraphics[scale=0.8]{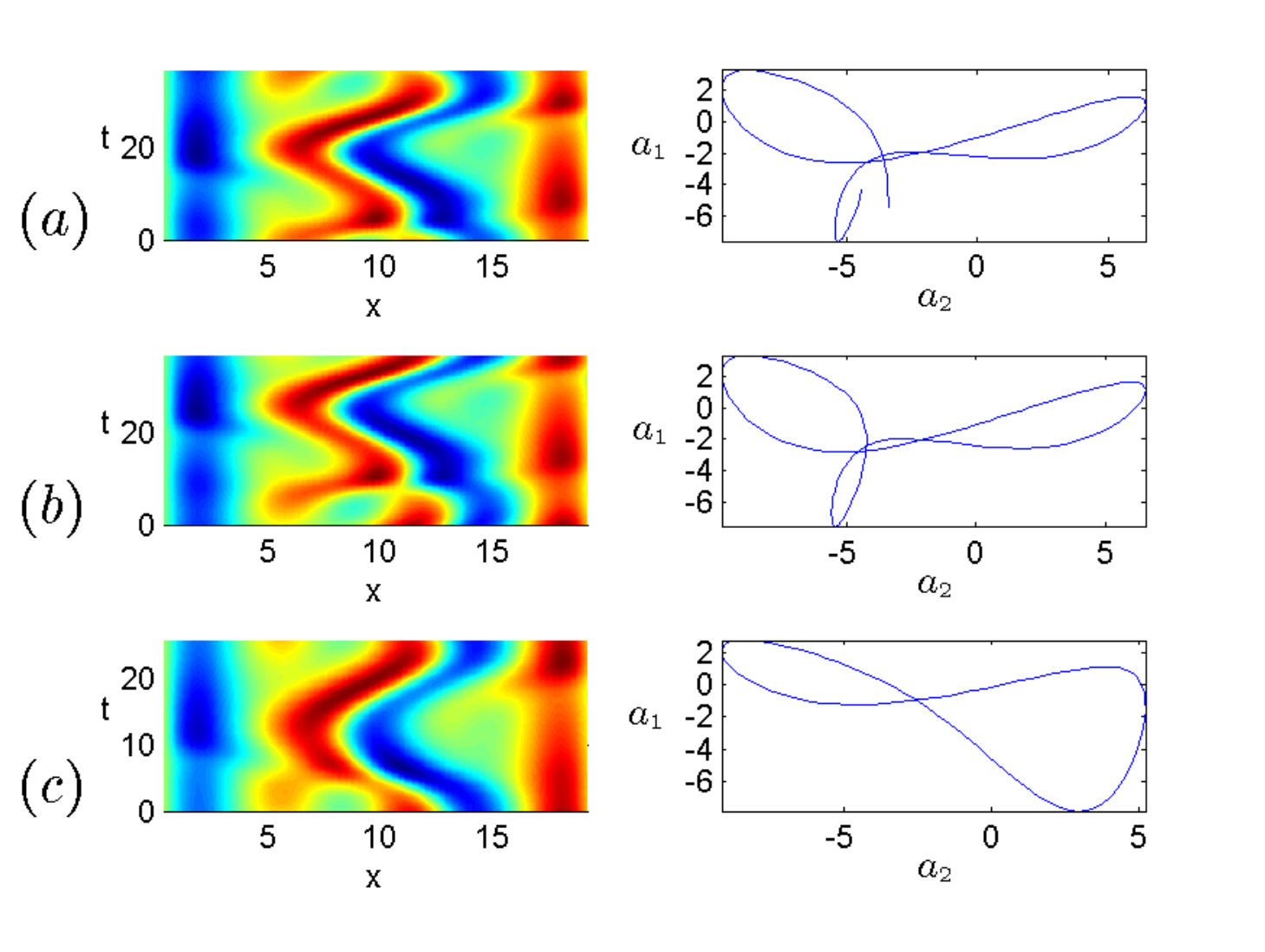}
\caption{Illustration of two UPOs of KSE detected from a single
seed. We show both a level plot for the solutions and a projection
onto the first two Fourier components. Since $u(x,t)$ is
antisymmetric on $[0,L]$, it is sufficient to display the space-time
evolution of $u(x,t)$ on the interval $[0, L/2]$: (a) Seed with time
$T = 37.0$, (b) a periodic solution of length $T = 36.9266$ detected
with stabilising transformation $\tilde{C} = +1$ and (c) a periodic
solution of length $T = 25.8489$ detected with stabilising
transformation $\tilde{C} = -1$.}\label{fig:kse}
\end{center}
\end{figure}

\begin{table}[t]
 \caption{The number of distinct periodic solutions for the
 Kuramoto-Sivashinsky equation detected by the method of stabilising
 transformations. Here $L = 51.4$ and $\alpha = 0.25$.}
    \begin{center}
        \begin{tabular}{|c|c||c|c|c|c|c|}
            \hline
            Period & C & $N_{\mathrm{po}}$ & $N_{\mathrm{hit}}$ & $N_{\mathrm{fev}}$
            & $N_{\mathrm{jev}}$ & Work\\
            \hline
            \multirow{9}{*}{$10 - 100$} & $C_0$ & $11$ & $366$ & $382$ & $16$ & $894$ \\
            &  $C_1$ & $1$ & $221$ & $410$ & $23$ & $1146$ \\
            &  $C_2$ & $7$ & $108$ & $456$ & $20$ & $1096$ \\
            &  $C_3$ & $2$ & $81$  & $357$ & $13$ & $773$ \\
            &  $C_4$ & $2$ & $157$ & $464$ & $26$ & $1296$ \\
            &  $C_5$ & $0$ & $171$ & $654$ & $35$ & $1774$ \\
            &  $C_6$ & $5$ & $174$ & $666$ & $36$ & $1818$ \\
            &  $C_7$ & $2$ & $139$ & $496$ & $36$ & $1648$ \\ \cline{3-7}
            &  $\{C_i\}$ & $30$ & $1417$ & $3885$ & $205$ & $10445$ \\ \hline
            \hline
            \multirow{9}{*}{$100 - 250$} & $C_0$ & $51$ & $330$ & $628$ & $17$ & $1172$ \\
            & $C_1$ & $7$  & $209$ & $807$ & $27$ & $1671$ \\
            & $C_2$ & $17$ & $138$ & $936$ & $31$ & $1928$ \\
            & $C_3$ & $21$ & $177$ & $917$ & $26$ & $1749$ \\
            & $C_4$ & $11$ & $161$ & $877$ & $36$ & $2029$ \\
            & $C_5$ & $0$  & $116$ & $960$ & $43$ & $2336$ \\
            & $C_6$ & $1$  & $117$ & $975$ & $42$ & $2319$ \\
            & $C_7$ & $6$  & $161$ & $879$ & $35$ & $1999$ \\ \cline{3-7}
            & $\{C_i\}$ & $114$ & $1409$ & $6979$ & $257$ & $15203$ \\
            \hline
        \end{tabular}
    \end{center}\label{table4}
\end{table}

\begin{table}
 \caption{The number of distinct periodic solutions for the Kuramoto-Sivashinsky
 equation detected by the Levernberg-Marquardt algorithm {\tt lmder} with $L = 51.4$}
    \begin{center}
        \begin{tabular}{|c|c||c|c|c|c|c|}
            \hline
            Method & Period & $N_{\mathrm{po}}$ & $N_{\mathrm{hit}}$ & $N_{\mathrm{fev}}$
            & $N_{\mathrm{jev}}$ & Work\\
            \hline
            \multirow{2}{*}{Constrained} & $10-100$ & $19$ & $311$ & $26$ & $18$ & $602$ \\
            &  $100-250$ & $85$ & $337$ & $72$ & $61$ & $2024$ \\ \hline
            \hline
            \multirow{2}{*}{Unconstrained} & $10-100$ & $19$ & $489$ & $16$ & $12$ & $400$ \\
            &  $100-250$ & $85$ & $396$ & $63$ & $51$ & $1695$ \\ \hline
        \end{tabular}
    \end{center}\label{table5}
\end{table}

\begin{table}[t]
 \caption{The number of distinct periodic solutions for the Kuramoto-Sivashinsky
 equation detected by the Newton-Armijo algorithm with $L = 51.4$.}
    \begin{center}
         \begin{tabular}{|c||c|c|c|c|c|}
            \hline
            Period & $N_{\mathrm{po}}$ & $N_{\mathrm{hit}}$ & $N_{\mathrm{fev}}$ & $N_{\mathrm{jev}}$ &
            Work\\\hline
            $10 - 100$  & $19$ & $308$ & $73$ & $20$ & $713$ \\
            $100 - 250$ & $75$ & $309$ & $174$ & $43$ & $1550$ \\
            \hline
        \end{tabular}
    \end{center}\label{table6}
\end{table}

Similarly, we run the {\tt lmder} and NA routines from the same $500$ seeds over the two different time ranges. The respective routines are terminated if one of the following three scenarios arise: (i) a predefined maximum number of function evaluations is exceeded, we set the maximum number of function evaluations equal to $1000$, (ii) the error between two consecutive steps is less than xtol, but the sum of squares is greater than ftol, indicating a local minimum has been detected, or (iii) both xtol and ftol are satisfied indicating that convergence to a UPO has been obtained.

Note that, one problem with applying methods that use a cost function to obtain ``global'' convergence to our setting, is that for increasing period, the level curves of $|g|^2$ become increasingly compressed along the unstable manifold of $\phi^T(x)$, resulting in a complicated surface with many minima, both local and global, embedded within long, winding, narrow ``troughs''.

This can be explained by the following heuristics: for simplicity let us assume that we are dealing with a map $x_{n+1} = f(x_n)$, whose unstable manifold is a one-dimensional object. In that case, we may define a one-dimensional  map, locally, about a period-p orbit, $x^*$, of the map $f$ as
\begin{equation*}\label{eqn:localmap}
    \bar{h}(s) = |g(x^*+\delta x)|^2.
\end{equation*}
Here $g = f^p(x)-x$ as usual, $\delta x = x(s) - x^*$ is small, and we only allow $x(s)$ to vary along the unstable manifold. Expanding $g$ in a Taylor series about the periodic orbit, $x^*$, we obtain
\begin{eqnarray*}\label{eqn:quad}
    \bar{h}(s) &=& |g(x^*) + Dg(x^*)\delta x + O(\delta x^2)|^2\\
               &=& |(Df^p(x^*)-I_n)\delta x + O(\delta x^2)|^2\\
               &=& \delta x^{\mathsf{T}}(\Lambda^p-I_n)^2\delta x + O(\delta x^3),
\end{eqnarray*}
where the third line follows since $\delta x$ is an eigenvector of $Df(x^*)$ with corresponding eigenvalue $\lambda$, and $\Lambda = \mathrm{diag}(\lambda,\dots\lambda)$. Note that in the above we assume that $p$ is large but finite so that the term $(\Lambda^p-I_n)^2$ remains bounded. Now, close to the periodic orbit, $\bar{h}$ is approximately a quadratic form with slope of the order $\lambda^p$, and since $|\lambda|>1$, it follows that if we move along the unstable manifold $|g|^2$ will grow quicker for larger periods, or, in other words, the level curves are compressed along the unstable direction.

Since both {\tt lmder} and NA reduce the norm of $g = \phi^T(x) - x$ at each step, they will typically follow the gradient to the bottom of the nearest trough, where they will start to move along the narrow base towards a minimum. Once at the base of a trough, however, both algorithms are forced into taking very small steps, this follows due to the nature of the troughs, i.e. the base is extremely narrow and winding, and since both methods choose there next step-size based on the straight line search. In order to avoid this situation we use the following additional stopping criteria in our experiments
\begin{equation}\label{eqn:stopping}
  N_{\mathrm{step}}  > \left\lceil \frac{||g(x_n)||}{||g(x_{n-1})|| - ||g(x_n)||}\right\rceil.
\end{equation}
Here $N_{\mathrm{step}}$ denotes the maximum number of iterations allowed, and the term on the right is a linear approximation of the number of steps required for convergence. If the above condition fails in $50$ consecutive iterates we terminate the search.

In order to make a comparison between the efficiency of the methods we introduce the measure of the work done per seed
\begin{equation}\label{eqn:cost}
    \mathrm{Work} = N_{\mathrm{fev}} + n\times N_{\mathrm{jev}}.
\end{equation}
Here $N_{\mathrm{fev}}$ is the average number of function evaluations per seed, $N_{\mathrm{jev}}$ the average number of Jacobian evaluations per seed and $n$ is the size of the system being solved. The expression in (\ref{eqn:cost}) takes
into account the fact that evaluation of the Jacobian is $n$ times more expensive than evaluation of the function itself.

The results of our experiments are summarised in Tables \ref{table1}--\ref{table6}. Here $N_{\mathrm{po}}$ denotes the number of distinct orbits found, $N_{\mathrm{hit}}$ gives the number of times we converged to a UPO, and $N_{\mathrm{fev}}$, $N_{\mathrm{jev}}$, and Work, are as defined above. In Tables \ref{table1} and \ref{table4} the performance of the stabilising transformations are analysed both collectively and on an individual basis; here the different $C_i$ denote the different SD matrices. In Table \ref{table1} $C_0 = +1$ and $C_1 = -1$, whilst in Table \ref{table4} we have
\begin{eqnarray*}
   \left\{
    C_{0}=\left[\begin{array}{cc}1&0\\0&1\\\end{array}\right],\right.
    C_{1}=\left[\begin{array}{cc}-1&0\\0&1\\\end{array}\right],
    C_{2}=\left[\begin{array}{cc}1&0\\0&-1\\\end{array}\right],
    C_{3}=\left[\begin{array}{cc}-1&0\\0&-1\\\end{array}\right],\\
    C_{4}=\left[\begin{array}{cc}0&1\\1&0\\\end{array}\right],
    C_{5}=\left[\begin{array}{cc}0&-1\\1&0\\\end{array}\right],
    C_{6}=\left[\begin{array}{cc}0&1\\-1&0\\\end{array}\right],\left.
    C_{7}=\left[\begin{array}{cc}0&-1\\-1&\\\end{array}\right]\right\}
\end{eqnarray*}
The total work done per seed is given by the sum over all the SD matrices and is denoted by $\{C_i\}$.

In total we found $487$ UPOs using the ST method, $379$ UPOs using {\tt lmder} and $350$ UPOs using NA. Whilst all methods found roughly the same number of orbits when searching for shorter period cycles, Tables \ref{table1} -- \ref{table4} reveal significant differences in performance. By comparing the work done per seed we see that both {\tt lmder} and NA are considerably faster, indeed efficiency between ST and the other methods for $L = 38.5$ differ by factors as great as $10$. The situation changes, however, when we look at the detection of longer cycles. For $L = 38.5$ in particular, we see that the ST method computes many more orbits than its competitors. Also, although both {\tt lmder} and NA remain more efficient than the ST method, the difference in efficiency is now only a factor of $3$ in the case of {\tt lmder} and $5$ in the case of NA. In fact, using the identity matrix alone, the ST method detects more UPOs than either {\tt lmder} or NA, but with comparable efficiency. For larger system size, $L = 51.4$, the ST method still detects more orbits than {\tt lmder} or NA. However, in doing so a considerable amount of extra work is done. It is important to note here, that the increase in work is due, mainly, to the fact that not all the SD matrices work well. For example, the subset of matrices $\{C_0, C_2, C_3, C_4\} \subset \mathcal{C}_{ \mathrm{SD}}$, detect approximately $90\%$ of the longer period UPOs, yet they account for less than $50\%$ of the overall work. Note that it is not surprising that the SD matrices do not perform equally well, this is, after all, what we would have expected based upon our experience with maps. However, it is still important to try all SD matrices -- when possible -- in order to compare their efficiency, especially if one wishes to construct minimal sets of ST matrices.

Another important consideration is, how was the performance of {\tt lmder} affected by the additional constraint? We can see from Tables \ref{table3} and \ref{table5} that the unconstrained system is more efficient, in all cases requiring fewer steps to converge, more importantly, it can be seen that the constrained system can fail to converge at all. Crucially, the number of searches which failed increases considerably as we look at larger system size, therefore, as we move to more complicated systems we would expect to see this difference in performance further increase. For example, in \cite{Lopez05} Lopez {\em et al} use {\tt lmder} to search for relative periodic orbits in the complex Ginzburg-Landau equation, where they have augmented the system with three additional equations. Our results suggest that this search would have benefited, not only in performance and numbers of UPOs detected, but by the savings in both time and effort required to construct additional equations and the resulting Jacobian, by setting all additional equations identically equal to zero.

Finally, a key feature of our method is that we can converge to several different UPOs from just one seed, depending upon which ST is used. Figure \ref{fig:kse} shows one of such cases, where Eq.~(\ref{eqn:augflow}) is solved from the same seed for each of the $2^{n_u}$ STs ($n_u = 1$ in this example), with two of them converging to two different UPOs. Figure \ref{fig:kse}a shows the level plot of the initial condition and a projection onto the first two Fourier components. Figures \ref{fig:kse}b and \ref{fig:kse}c show two unstable spatiotemporally periodic solutions which where
detected from this initial condition, the first of period $T = 36.9266$ was detected using $\tilde{C} = +1$, whilst the second of period $T = 25.8489$ was detected using $\tilde{C} = -1$. The ability to detect several orbits from one seed increases the efficiency of the algorithm.

\subsubsection{Seeding with UPOs}
In order to construct a seed from an already detected orbit, $(a^*, T^*)$, we begin by searching for close returns. That is, starting from a point on the orbit, $a^*(t_0)$, we search for a time $t_1$ such that $a(t_1)$ is close to $a(t_0)$. As long as $\tilde{T}\mathrm{mod}T\neq 0$, where $\tilde{T} = t_1 - t_0$, we take $(a^*(t_0), \tilde{T})$ as our initial guess to a time periodic solution. For longer period cycles, we can find many close returns by integrating just once over the period of the orbit. Shorter cycles, however, produce fewer recurrences and, in general, must be integrated over longer times to find good initial seeds. In our experiments we searched for close returns $\tilde{T}\in(10, 2T^*)$ if $T^*<150.0$; otherwise, we chose $\tilde{T}\in(10, T^*)$.

Stabilising transformations are constructed by applying the polar decomposition to the matrix $\tilde{G} = \tilde{Q}\tilde{B}$, where $\tilde{G}$ is defined by
\begin{equation*}
   \tilde{G} := V(S\Lambda - \mathrm{I}_n)V^{-1}.
\end{equation*}
Here $V$ and $\Lambda$ are defined through the eigen decomposition of $D\phi^{T^*}(a^*(t_0)) = V\Lambda V^{-1}$, and $S = \mathrm{diag}(\pm 1, \pm 1,\ldots,\pm 1)$.  The different transformations are then given by $C = - \tilde{Q}^{\mathsf{T}}$. Now, in~\cite{Crofts06} it was shown that a change in the signs of the stable eigenvalues does not result in a substantially different stabilising transformation, thus we use the subset of $S$ such that $S_{ii} = 1$ for $i>n_u$. For $L = 38.5$, we have just two such transformations, since all periodic orbits have only one unstable eigenvalue. Whilst for $L=51.4$, we will have either two or four transformations depending upon the number of unstable eigenvalues of $D\phi^{T^*}(a^*(t_0))$.

In the calculations to follow, a close return was accepted if $|a^*(t_1) - a^*(t_0)|<2.5$. Note that, this was the smallest value to produce a sufficient number of recurrences to initiate the search. A particular cycle may exhibit many close returns. In the following, we set the maximum number of seeds per orbit equal to $5$. To obtain convergence within tolerance $10^{-7}$ we had to increase the integration time to $s = 250$ and the maximum number of integration steps allowed to $2000$. If we converged to a UPO then as in the preceding section, we apply two or three Newton-Armijo steps to the Eq.~(\ref{eqn:augsys}) in order to converge to machine precision.

\begin{table}[t]
 \caption{Number of distinct orbits detected using the method of
 stabilising transformations with periodic orbits as seeds The number
 of seeds is $489$. $L = 38.5$, $\alpha = 0.25$.}
    \begin{center}
        \begin{tabular}{|c||c|c|c|c|c|}
            \hline
            C & $N_{\mathrm{po}}$ & $N_{\mathrm{hit}}$ & $N_{\mathrm{fev}}$ & $N_{\mathrm{jev}}$ &
            Work\\\hline
            $C_0$ & $46$ & $209$ & $2814$ & $136$ & $4854$ \\
            $C_1$ & $52$ & $198$ & $2819$ & $130$ & $4769$ \\\cline{2-6}
            $\{C_i\}$ & $98$ & $407$ & $5633$ & $266$ & $9623$ \\
            \hline
        \end{tabular}
    \end{center}\label{table7}
\end{table}

\begin{table}[t]
 \caption{Number of distinct orbits detected using the method of
 stabilising transformations with periodic orbits as seeds. The number
 of seeds is $123$. $L = 51.4$, $\alpha = 0.25$.}
    \begin{center}
        \begin{tabular}{|c||c|c|c|c|c|}
            \hline
            C & $N_{\mathrm{po}}$ & $N_{\mathrm{hit}}$ & $N_{\mathrm{fev}}$ & $N_{\mathrm{jev}}$ &
            Work\\\hline
            $C_0$ & $1$ & $21$ & $2797$ & $48$ & $4333$ \\
            $C_1$ & $2$ & $37$ & $2815$ & $40$ & $4095$ \\
            $C_2$ & $0$ & $0$ & $216$ & $3$ & $312$ \\
            $C_3$ & $0$ & $1$ & $187$ & $2$ & $251$ \\\cline{2-6}
            $\{C_i\}$ & $3$ & $59$ & $6015$ & $93$ & $8691$ \\
            \hline
        \end{tabular}
    \end{center}\label{table8}
\end{table}

The results of our experiments are given in Tables \ref{table7} and\ref{table8}. As in the previous section,
$N_{\mathrm{po}}$ denotes the number of distinct orbits found, $N_{\mathrm{hit}}$ the number of times we converged
to a UPO, $N_{\mathrm{fev}}$ the average number of function evaluations per seed, and $N_{\mathrm{jac}}$ the average
number of jacobian evaluations per seed. The computational cost per seed is measured in terms of the average number
of function evaluations per seed and is defined as in Eq.~(\ref{eqn:cost}). In Tables \ref{table7} and \ref{table8}
the different $C_i$ can be uniquely identified by the signature of pluses and minuses defined through the corresponding
matrix $S$. For $L = 38.5$, the matrices $C_0$ and $C_1$ correspond to the signatures $(+,\ldots,+)$ and $(-,+,\ldots,+)$, respectively, whilst for $L = 51.4$, the matrices $C_0$, $C_1$, $C_2$ and $C_3$ correspond to $(+,\ldots,+)$, $(-,+,\ldots,+)$, $(-,-,+,\ldots,+)$, and $(+,-,\ldots,+)$ respectively. As before, the total work done per seed is defined as the sum over all matrices and is denoted by $\{C_i\}$.

\begin{figure}[t]
\begin{center}
\includegraphics[scale=0.8]{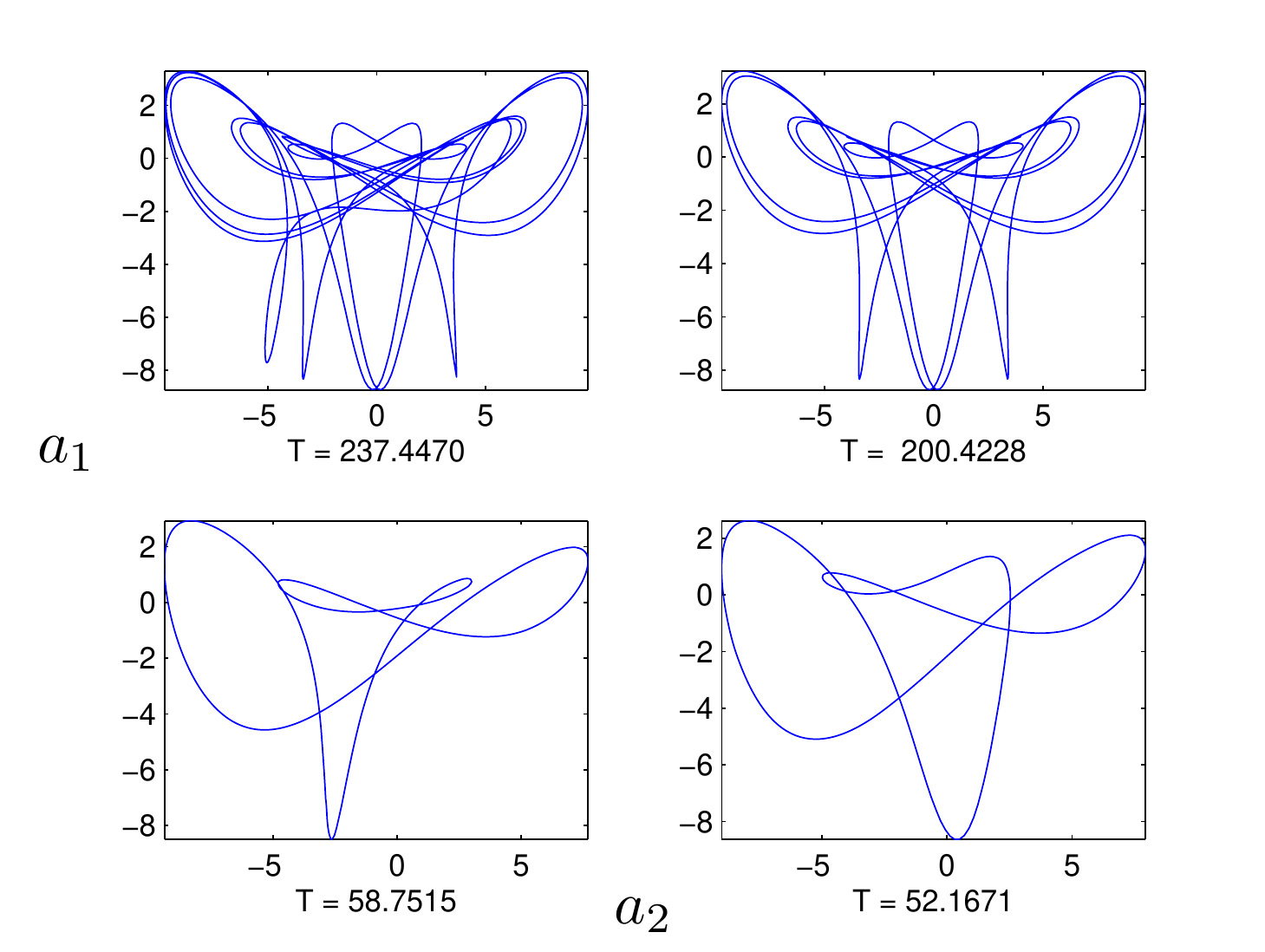}
\caption{Illustration of how a UPO can be used as a seed to detect
 new cycles. We show the projection onto the first two Fourier
 components of the initial seed ($T = 237.4470$) and the three detected
 orbits. Here $L = 38.5$ and $n = 15$.}\label{fig:seed}
\end{center}
\end{figure}

For $L = 38.5$ we where able to construct $489$ seeds from $343$ previously detected periodic orbits. From the new seeds we found a further $98$ distinct UPOs, bringing the total number of distinct orbits detected for $L = 38.5$ to $433$. An important observation to make is that both matrices perform equally well, as can be seen from Table \ref{table7}. This is in contrast to the SD matrices for which the performance of the different matrices varies greatly; see Tables \ref{table1} and \ref{table4}. The result of using periodic orbits as seeds has, however, increased the cost per seed as compared with the cost of seeding with close returns within a chaotic orbit, see Tables \ref{table1} and \ref{table7}. The increased computation is due mainly to the fact that the close returns obtained from periodic orbits are not as good as those obtained from a chaotic orbit. Recall that the stabilising transformations are based on the local invariant directions of the orbit, and we would expect the performance to suffer as we move further from the seed.

For larger system size, where the system becomes more chaotic, the construction of seeds from close returns becomes
increasingly difficult. For $L = 51.4$, we constructed $123$ initial seeds from the $144$ periodic orbits using the method of near recurrences. Here we detected only $3$ new distinct UPOs. It is important to note, however, that although we do not find many new UPOs for the case $L = 51.4$, the method is still converging for approximately $20\%$ of all initial seeds. Also, in Table \ref{table8}, the poor performance of the matrices $C_2$, $C_3$ as compared to that of $C_0$, $C_1$, is due to the fact that only a small percentage of seeds were constructed from orbits with two positive Lyapunov exponents.

One advantage of using periodic orbits as seeds is that we can construct many seeds from a single orbit. Figure
\ref{fig:seed} gives an illustration of three periodic orbits which were detected from a long periodic orbit used as seed. Figure \ref{fig:seed} shows the projection onto the first two Fourier components of the initial seed (of period $T = 237.4470$) and the three orbits detected whose periods in descending order are $T = 200.428$, $T = 58.7515$ and $T =
52.1671$. For each seed determined from a particular periodic orbit the stability transformations are the same. Hence,
the ability to construct several seeds from a single orbit, increases the efficiency of the scheme.

\section{Conclusions and further work}
We have presented a scheme for detecting UPOs in high dimensional chaotic systems based upon the stabilising transformations proposed in~\cite{Crofts06,Lai99,Schmelcher97}. Due to the fact that one often wishes to study low dimensional dynamics embedded in a high dimensional phase space, it is possible to increase the efficiency of the stabilising transformations approach by restricting the construction of such transformations only to the low-dimensional unstable subspace. Following the approach often adopted in subspace iteration methods~\cite{Lust98}, we construct a decomposition of the tangent space into unstable and stable orthogonal subspaces.  We find that the use of SVD to approximate the appropriate subspaces is preferable to that of Schur decomposition, which is usually employed within the subspace iteration approach. As illustrated with the example of the Ikeda map, the decomposition based on SVD is less susceptible to variations in the properties of the tangent space away from a seed and thus produce larger basins of attraction for stabilised periodic orbits. Within the low-dimensional unstable subspace, the number of useful stabilising transformations is relatively small, so it is possible to apply the full set of SD matrices.  In fact, we have found that the subset of diagonal matrices of $\pm 1$ is capable of locating a large number of UPOs, although more analysis will be carried out in the future to determine if it is possible to detect all types of UPOs with this subset.

In conclusion, we have presented an extension of the stabilising transformations approach for locating periodic orbits in high-dimensional dynamical systems. Future work will concentrate on rigorous mathematical analysis of this approach in order to determine the range of its applicability.  We will also work on the development, within the stabilising transformations approach, of an efficient strategy for systematic detection of periodic orbits in high-dimensional systems.


\begin{thebibliography}{30}

\bibitem{Benettin80}
{\sc G.~Benettin and L.~Galgani and A.~Giorgilli and
J.~M.~Strelcyn}, {\em Lyapunov characteristic exponents for smooth
dynamical systems and for {H}amiltonian systems; a method for
computing all of them. {P}art 2: Numerical application}, Meccanica
\textbf{15} pp.~21--29,(1980).

\bibitem{Crofts06}
{\sc J.~J. Crofts, R.~L. Davidchack}, {\em Efficient Detection of
Periodic Orbits in Chaotic Systems by Stabilising Transformations},
SIAM J.~Sci.~Comput. \textbf{28} 4 pp.~1275--1288 (2006).

\bibitem{Cvitanovic03}
{\sc P.~Cvitanovi\'c, Y.~Lan}, in {\em Proceedings of the 10th
International Workshop on Multiparticle Production: Correlations and
Fluctuations in QCD}, edited by {\sc N. ~Antoniou}, World
Scientific, Singapore, (2003).

\bibitem{Christiansen97}
{\sc F. Christiansen, P. Cvitanovi�c, and V. Putkaradze}, {\em
Spatiotemporal chaos in terms of unstable recurrent patterns},
Nonlinearity \textbf{10} pp.~55, (1997).

\bibitem{Lai99}
{\sc R.~L. Davidchack, Y.~C Lai}, Phys. Rev. E \textbf{60}
pp.~6172--6175 (1999).

\bibitem{HairerBook}
{\sc E. Hairer, G. Wanner}, {\em Solving Ordinary Differential
Equations II: Stiff and Differential-Algebraic Problems}, Springer,
New York, 2nd ed., (1996).

\bibitem{ODEPACK}
{\sc A.~C.~Hindmarsh}, {\em ODEPACK, A  systemized collection of ODE
solvers}, Sci.~Comput. (1983)

\bibitem{Ikeda79}
{\sc K. Ikeda}, Opt. Commun. \textbf{30} 257 (1979); {\sc S.~M.
Hammel, C.~K.~R.~T Jones, and J. Maloney}, J. Opt. Soc. Am B
\textbf{2} pp.~552 (1985).

\bibitem{Kassam05}
{\sc A.~Kassam, L.N.~Trefethen}, {\em Fourth Order Time Stepping for
Stiff PDEs}, SIAM J. Sci. Comput. \textbf{26} 4 pp.~1214--1233
(2005).

\bibitem{KelleyBook}
{\sc C.~T.~Kelley}, {\em Solving Nonlinear Equations with Newton's Method},
 Fundamentals of Algorithms. SIAM, 1st edition (2003).

\bibitem{Kevrekidis90}
{\sc I.G.~Kevrekidis, B. Nicolaenko and J. Scovel}, {\em Back in the
saddle again: a computer assisted study of the Kuramoto-Sivashinsky
equation}, SIAM J. Appl. Math \textbf{50} pp.~760 (1990).

\bibitem{Kuramoto76}
{\sc Y.~Kuramoto, T.~Tsuzuki}, Prog. Theor. Phys. \textbf{55} 365
(1976).

\bibitem{Lan04}
{\sc Y.~Lan, P.~Cvitanovi\'c}, Phys. Rev. E \textbf{69} pp.~016217
(2004).

\bibitem{Lopez05}
{\sc V.~Lopez and P.~Boyland and M.~T.~Heath and R.~D.~Moser}, {\em
Relative periodic solutions of the complex {G}inzburg-{L}andau
equation}, SIAM J.~Appl.~Dynamical~Systems \textbf{4} 4
pp.~1042--1075 (2005).

\bibitem{Lust98}
{\sc K.~Lust, D.~Roose, A.~Spence, A.~R. Champneys}, SIAM J. Sci.
Comput. \textbf{19} pp.~1188--1209 (1998).

\bibitem{Marquardt63}
{\sc D.~Marquardt} {\em An algorithm for least-squares estimation of
nonlinear parameters}, SIAM J.~Appl.~Math textbf{11} pp.~431--441
(1963).

\bibitem{Minpack}
{\sc J.~J.~Mor\'{e} and B.~S.~Gorbow and K.~E.~Hillstrom}, {\em User
guide for {\tt MINPACK-1}}, Technical report ANL-80-74, Argonne
national laboratory, {\tt http://www.netlib.org/minpack/}, (1980).

\bibitem{OttBook}
{\sc E.~Ott}, {\em Chaos in Dynamical Systems}, Cambridge University
Press, Cambridge, (1993).

\bibitem{Schmelcher97}
{\sc P.~Schmelcher and F.~K. Diakonos}, {\em Detecting unstable
periodic orbits of chaotic dynamical systems}, Phys. Rev. Lett.
\textbf{78} pp.~4733--4736 (1997).

\bibitem{Schmelcher98}
\leavevmode\vrule height 2pt depth -1.6pt width 23pt, {\em General
approach to the localization of unstable periodic orbits in chaotic
dynamical systems}, Phys. Rev. E \textbf{57} pp.~2739--2746 (1998).

\bibitem{Gear79}
{\sc L.~F.~Shampine and C.~W.~Gear}, {\em A user's view of solving
stiff ordinary differential equations }, SIAM Review \textbf{21} 1
pp.~1--17 (1979).

\bibitem{Keller93}
{\sc G.~M. Shroff, H.~B. Keller}, SIAM J. Numer. Anal \textbf{30}
pp.~1099--1120 (1993).

\bibitem{Sivashinsky77}
{\sc G.~I. Sivashinsky}, Acta Astron. \textbf{4} pp.~1177 (1977).

\bibitem{TrefethenBook}
{\sc L.~N. Trefethen}, {\em Spectral Methods in Matlab}, SIAM,
Philadelphia, (2000).

\bibitem{NumericalRecipes}
{\sc William~H. Press, Saul~A. Teukolsky, William~T. Vetterling and
Brian~P. Flannery}, {\em Numerical Recipes in C}, Cambridge
University Press, cambridge, (1992)

\bibitem{Zoldi98}
{\sc S.~M. Zoldi, H.~S. Greenside}, Phys. Rev. E \textbf{57}
pp.~2511--2514 (1998).

\end{thebibliography}
\end{document}